\documentclass[twocolumn,showpacs,superscriptaddress,pre]{revtex4-1}
\usepackage{amsmath}
\usepackage{graphicx}
\usepackage{bm}
\usepackage{epstopdf}
\usepackage{caption}
\usepackage{subcaption}

\DeclareMathOperator{\erf}{erf}

\DeclareMathOperator{\Ei}{E_i}

\newcommand{\be}{\begin{equation}}
\newcommand{\ee}{\end{equation}}

\begin{document}

\title{Variational approach to multidimensional solitons in highly
nonlocal nonlinear media}
\author{Branislav N. Aleksi\'c}
\affiliation{Institute of Physics, University of Belgrade, P.O.Box 68, 11080 Belgrade,
Serbia}
\affiliation{Texas A\&M University at Qatar, P.O.Box 23874, Doha, Qatar}
\author{Najdan B. Aleksi\'c}
\affiliation{Institute of Physics, University of Belgrade, P.O.Box 68, 11080 Belgrade,
Serbia}
\affiliation{Texas A\&M University at Qatar, P.O.Box 23874, Doha, Qatar}
\author{Milan S. Petrovi\'c}
\affiliation{Institute of Physics, P.O.Box 57, 11001 Belgrade, Serbia}
\author{Aleksandra I. Strini\'c}
\affiliation{Institute of Physics, University of Belgrade, P.O.Box 68, 11080 Belgrade,
Serbia}
\author{Milivoj R. Beli\'c}
\affiliation{Texas A\&M University at Qatar, P.O.Box 23874, Doha, Qatar}

\begin{abstract}
We apply the variational approach to solitons in highly nonlocal nonlinear media in $D=1,2,3$ dimensions.
We compare results obtained by the variational approach with those obtained in the accessible
soliton approximation, by considering the same system of equations
in the same spatial region and under the same boundary conditions.
To assess the accuracy of these approximations, we also compare them with the numerical solution of the equations.
We discover that the variational highly nonlocal approximation provides more accurate
results and as such is more appropriate solution than the accessible soliton approximation.

\end{abstract}

\maketitle


\section{Introduction}

Optical spatial
solitons are self-localized wave packets that propagate in a nonlinear medium
without changing their structure \cite{yuri}. This is made possible by the robust
balance between dispersion and nonlinearity or between
diffraction and nonlinearity or between all three effects in the
propagation of spatiotemporal localized optical fields.
An important characteristic of real nonlinear media is their nonlocality,
that is, the fact that characteristic size of
the response of the medium is wider than the size of the excitation
itself. Strong nonlocality is of special interest, because it is observed
in many media. For example, in nematic liquid crystals (NLCs) both
experimental and theoretical studies indicated that the 
nonlinearity is highly nonlocal -- meaning that the size of response is much
wider than the size of excitation \cite{hen,cyril,beec}.

In 1997, Snyder and Mitchell introduced a model of nonlinearity
whose response is highly nonlocal \cite{snyder} -- in fact,
infinitely nonlocal. They proposed an
elegant theoretical model, intimately connected with the linear
harmonic oscillator, that described complex soliton dynamics
in simple terms, even in two and three dimensions. Because of the simplicity of 
the theory, they coined the term "accessible solitons" (ASs) for these optical
spatial solitary waves. However, straightforward application of the AS
theory, even in nonlinear media with almost infinite range
of nonlocality, inevitably led to additional problems
\cite{assanto2003,assanto2004,henninot}, because there exists no real
physical medium without boundaries and without noise.

To include interactions between solitons within
boundaries, as well the impact of the finite size of the sample,
we developed a variational approach (VA) to solitons in
nonlinear media  with long-range nonlocality, such as NLCs \cite{aleksic},
materials with thermal nonlocality \cite{bucco},
photo-refractive crystals \cite{belic}, and Bose-Einstein condensates \cite{bec}.
Starting from an appropriate ansatz, this approach delivers a stationary solution
for the beam amplitude and width, as well as the period of small
oscillations about the stationary state. It provides for natural explanation
of oscillations seen when, e.g., noise is included into the nonlocal nonlinear models.
The noise is inevitable in any real physical system and causes a
regular oscillation of soliton parameters with the period well predicted
by our VA calculus \cite{aleksic}. It may even destroy solitons \cite{petrovic}. Although our
VA results were corroborated by numerics and experiments, they still attracted
a heated exchange with other researchers \cite{comment,reply}.
We have investigated in detail the destructive
influence of noise on the shape-invariant solitons in a highly nonlocal
NLCs in \cite{petrovic}.

Owing to great impact and practical relevance
of the AS model, in this paper we systematically compare it with the VA approximation in
highly nonlocal nonlinear media in $D = 1, 2, 3$ dimensions. We
consider both systems using the same general equations, in the same
spatial region, and under the same boundary conditions. To check the
accuracy of both approximations, we compare them with the numerical
solution of the equations. We find that multidimensional variational highly nonlocal
approximation provides very accurate results, while the beam parameters
obtained by using the AS approximation show systematic and predictable
discrepancy with numerical results.

The paper is organized in the following manner. In Sec. 2 we introduce the model,
Sec. 3 presents results obtained by the variational approach, Sec. 4 discusses
the accessible soliton approximation, Sec. 5 presents numerical results, and
Sec. 6 brings conclusions.

\section{The model}

For the study of VA and AS approximations to the fundamental
soliton solutions in a ($D$+1)-dimensional highly nonlocal medium we
adopt the following model of coupled dimensionless equations \cite{yuri,aleksic}:

\begin{equation}
2i{\frac{\partial E}{\partial z}}+\Delta E+\theta E=0,  \label{eq1}
\end{equation}

\begin{equation}
2\Delta \theta -\alpha \theta +\left\vert E\right\vert ^{2}=0,
\label{eq2}
\end{equation}

\noindent with zero boundary conditions on a $D$-dimensional sphere $\left( D=1,2,3\right)$.
Here $\Delta$ is the $D$%
-dimensional Laplacian. The system of equations of interest consists of the nonlinear
Schr\"{o}dinger equation for the propagation of the optical field $E$ and
the diffusion equation for the nonlocal response of the medium $\theta $.
This is a fairly general model for nonlinear optical media with diffusive
nonlocality, widely used in the literature \cite{yuri,assanto2003,aleksic}.
In the local limit, the first term in Eq. (\ref{eq2}) can be neglected and
the model reduces to the Schr\"{o}dinger equation with Kerr nonlinearity. In
the opposite limit, the second term in Eq. (\ref{eq2}) can be neglected and
the highly nonlocal model is reached. Since we are interested in the strong
nonlocality, we will omit in our analysis the $\alpha $ term from Eq. (\ref{eq2}).

For radially-symmetric
intensity distributions $\left\vert E\right\vert ^{2}$, Eq. (\ref{eq2}) can be solved
using Green's functions,

\begin{equation}
\theta _{D}(r)=\frac{1}{2}\int G_{D}(r,\rho )\left\vert E(\rho )\right\vert ^{2}\rho
^{D-1}d\rho ,  \label{sol2gen}
\end{equation}%
where $G_{D}(r,\rho )$ \ is the Green's function in $D$ dimensions,
defined as follows:

\begin{equation}
G_{D}(r,\rho )=\left\{
\begin{array}{c}
\left( r^{2-D}-d^{2-D}\right) /(D-2),\qquad \rho <r<d \\
\left( \rho ^{2-D}-d^{2-D}\right) /(D-2),\qquad r<\rho <d%
\end{array}%
\right\}.
\end{equation}%
Here $d$ is some characteristic transverse distance of interest.
In the two-dimensional case (here and thereafter), the limit $D\rightarrow
2 $ should be taken, leading to

\begin{equation}
G_{2}(r,\rho )=\left\{
\begin{array}{c}
\ln \left( r/d\right) ,\qquad \rho <r<d \\
\ln \left( \rho /d\right) ,\qquad r<\rho <d%
\end{array}%
\right\} .
\end{equation}%
For the Gaussian-shaped beams with the field intensity:
\begin{equation}
\left\vert E(r)\right\vert ^{2}=\frac{Q_D}{T^{D}\pi ^{D/2}}\exp \left( -\frac{%
r^{2}}{T^{2}}\right) ,
\end{equation}%
\noindent the solution of
Eq. (\ref{sol2gen}) can be written as


\begin{equation}
\begin{split}
\theta_D (r) =\frac{Q_D}{4\pi ^{D/2}(2-D)}
 \left\{r^{2-D}\Gamma (D/2,r^{2}/T^{2})- \right. \\
 \left.  T^{2-D}\exp (-r^{2}/T^{2})+   \left[ d^{2-D}-r ^{2-D}\right] \Gamma
(D/2)\right\} +O(\delta ),
\end{split}
\label{theta}
\end{equation}

\noindent where $\delta =\max \{\exp (-d^{2}/T^{2})\}\ll 1$ and $T$ is the characteristic width
of the optical field. The $D-$dimensional power is given by $%
Q_{D}=\nu \int \left\vert E\right\vert ^{2}r^{D-1}dr$, where $\nu =2\pi
^{D/2}/\Gamma ^{{}}(D/2)$.

In an explicit form for different dimensions, Eq. (\ref{theta}) yields:

\begin{equation}
\theta_{1} (r)=\frac{Q_1}{4}\left[d- r \erf\left(\frac{r}{T}\right) -\frac{T\exp(-r^2/T^2)}{\sqrt{\pi}} \right]  +O(\delta ),
\end{equation}

\begin{equation}
\theta_{2} (r)=\frac{Q_2}{8\pi}\left[ \Ei\left(-\frac{r^{2}}{T^{2}} \right)  -\ln \left(\frac{r^{2}}{d^{2}}\right) \right] + O(\delta),
\end{equation}

\begin{equation}
\theta_{3} (r)=\frac{Q_3}{8\pi}\left[\frac{\erf\left(r/T\right)}{r} -\frac{1}{d} \right]+O(\delta),
\end{equation}
\noindent for one, two, and three dimensions, respectively.

In the equations above, $\erf(z)=\left( 2/\sqrt{\pi }\right)
\int_{0}^{z}\exp (-t^{2})dt$ is the error function, $Ei=-\int_{-z}^{\infty }t^{-1}\exp (-t)dt$ is the exponential integral
function, $\Gamma (z)=\int_{0}^{\infty }t^{z}\exp (-t)dt$ is the Euler gamma
function, and $\Gamma (a,z)=\int_{z}^{\infty }t^{a-1}\exp (-t)dt$ is the
incomplete gamma function \cite{abramowitz}. The form of $\theta $ corresponds to a radially-symmetric solution of Eq.
(\ref{eq2}), with zero boundary conditions on a circle of radius $d\gg T$, and
in the limit of a thick transverse width. We can take these expressions as an approximate
solution on a $D-$dimensional cube, as demonstrated in \cite{aleksic,VAvsAS} for the two-dimensional case.

In the AS approximation, for the shape of the
nonlocal response $\theta (r)$ of the medium one uses a parabolic function of
the transverse distance. Expanding the solution of Eq. (\ref{theta}) in Taylor
series up to the second terms, one finds

\begin{equation}
\theta_D (r)\approx \theta _{D\max}-\Theta _{D}r^{2},
\label{parabolTheta}
\end{equation}

\noindent where the parabolicity coefficient is

\begin{equation}
\Theta _{D}=\frac{Q_D}{4D\pi ^{D/2}T^{D}},
\label{theta_D}
\end{equation}

\noindent and the maximum value of $\theta $ is

\begin{equation}
\theta _{D\max}=\frac{Q_D}{4\pi ^{D/2}}\frac{ d^{2-D}\Gamma (D/2)-
T^{2-D}}{2-D}+O(\delta).
\label{stac2}
\end{equation}

\noindent In two dimensions this becomes:

\begin{equation}
\theta _{2\max}=\frac{Q_2}{8\pi }\ln \left(\frac{ e^{\gamma }d^{2}}{T^{2}}\right)+O(\delta),
\end{equation}
which agrees with the value determined in \cite{aleksic,VAvsAS}.
Here $\gamma$ is Euler's constant.

\section{Variational Approach}

In the variational approach, to derive equations describing evolution of the 
field beam expressed in an appropriate approximate form, one introduces a Lagrangian density, corresponding to Eq. (%
\ref{eq1}):

\begin{equation}
\mathcal{L}_{D}=i\left( {\frac{\partial E^{\ast }}{\partial z}}E-{\frac{%
\partial E}{\partial z}}E^{\ast }\right) +\left\vert \nabla E\right\vert
^{2}-\theta_D \left\vert E\right\vert ^{2}.\qquad  \label{lagrang}
\end{equation}

\noindent Thus, the problem is reformulated into a variational problem

\begin{equation}
\delta \iint \nu \mathcal{L}_{D}r^{D-1}drdz=0,
\end{equation}

\noindent whose solution is equivalent to Eq. (\ref{eq1}).
To obtain evolution equation for an approximate field in the highly nonlocal
region, an ansatz is introduced in the form of a Gaussian beam for the field \cite{aleksic}:

\begin{equation}
E=A\exp \left[-{\frac{r^2}{2 R^2}}+iCr^2+i\psi \right] ,  \label{trial1}
\end{equation}

\noindent in which $A$ is the amplitude, $R$ is the beam width, $C$ is the
wave front curvature along the transverse coordinate, and $\psi $ is the
phase shift. Variational optimization of these beam parameters will lead to
the most appropriate VA solution of the problem. Likewise, a trial function
for the nonlocal response of the medium is introduced, in the form Eq. (\ref%
{theta}) which is characterized by the power $Q_D=P_D$ and the
thickness $T=R$ of the beam. Again, the form of $\theta$ corresponds to a
radially-symmetric solution of Eq. (\ref{eq2}), with zero boundary
conditions on a $D-$dimensional sphere of radius $d\gg R$ (the limit of a
thick cell $\delta \ll 1$).

The averaged Lagrangian $L_{D}=\nu \int \mathcal{L}_{D}r^{D-1}dr$ is given by:

\begin{equation}
\begin{split}
L_{D}=& 2P_{D}\psi ^{\prime }+DP_{D}R^{2}\left( C^{\prime }+2C^{2}\right)  + \\& +
\frac{DP_{D}}{2R^{2}}+U_{D}(Q_D,T,P_D,R)
\end{split}
\label{lagr}
\end{equation}

\noindent where the prime in Eq. (\ref{lagr}) denotes the derivative with respect to $z$ and

\begin{equation}
\begin{split}
& U_{D}=U_{D}(Q_D,T,P_D,R)=-\nu \int \vartheta _{D}\left\vert E\right\vert
^{2}r^{D-1}dr=
\\& =\frac{Q_DP_D}{4(2-D)\pi ^{D/2}}\left( \left( R^{2}+T^{2}\right) ^{\left(
2-D\right) /2}-d^{2-D}\Gamma (D/2)\right)+O(\delta).
\end{split}
\label{U_D}
\end{equation}

\noindent Specifically for $D=2$:

\begin{equation}
U_{2}=\frac{Q_2P_2}{8\pi} \ln \frac{R^{2}+T^{2}}{e^\gamma d^{2}} +O(\delta).
\end{equation}%
\noindent  In the process
of optimization from the averaged Lagrangian, one obtains four ordinary differential equations (ODEs)
for the beam parameters:

\begin{equation}
\frac{dP_{D}}{dz}=0,  \label{eul1}
\end{equation}

\begin{equation}
C=\frac{1}{2R}\frac{dR}{dz},  \label{eul2}
\end{equation}%
\begin{equation}
\frac{d^{2}R}{dz^{2}}=\frac{1}{R^{3}}-\frac{1}{DP_{D}}\left( \frac{\partial
U_{D}}{\partial R}\right) _{T=R,\ Q=P},  \label{eul3}
\end{equation}

\begin{equation}
\frac{d\psi }{dz}=-\frac{D}{2R^{2}}+\frac{1}{2}\left( \frac{R}{2P}\frac{%
\partial U_{D}}{\partial R}-\frac{\partial U_{D}}{\partial P}\right) _{T=R,\
Q=P},
\label{eul4}
\end{equation}

\noindent According to Eq. (\ref{eul1}), the beam power $P_D=\nu
\int |E|^{2}r^{D-1}dr=\pi ^{D/2}A^{2}R^{D}$ is conserved.
The system of Eqs. (\ref{eul1}-\ref{eul3}) describes the
dynamics of the beam around a stationary state.

In the stationary state $(dR/dz=dC/dz=C=0)$, we find the equilibrium beam
width $R$:

\begin{equation}
R_{VA}=\left(\frac{2^{2+D/2}\pi ^{D/2}D}{P_D}\right) ^{1/\left( 4-D\right)
}+O(\delta),
\label{stacRva}
\end{equation}

\noindent and the amplitude $A$:

\begin{equation}
\begin{split}
A_{VA}&=\frac{2^{1+D/4}\sqrt{D}}{R_{VA}^{2}}= \\& =\left( \frac{P_D^2}{2^{D(1+D/4)}D^{D/2}\pi^D} \right)^{1/(4-D)}+O(\delta).
\end{split}
\label{Ava}
\end{equation}

\noindent as functions of the beam power.

The period of small oscillations of the perturbation around the equilibrium
position ($C=0$) is given by the following relation:

\begin{equation}
\begin{split}
\Lambda _{VA}&=\frac{2\pi }{\sqrt{4-D}}R_{VA}^{2}=\\&
=\frac{2\pi }{\sqrt{4-D}}\left(\frac{2^{2+D/2}\pi ^{D/2}D}{P_D}\right) ^{2/\left( 4-D\right)
}+O(\delta). 
\end{split}
\label{period VA}
\end{equation}%

\noindent From relations (\ref{U_D}, \ref{eul4}, \ref{stacRva}) we also find that the
propagation constant $\mu = d\psi /dz$, in the stationary state, can be written as:

\begin{eqnarray}
\mu_D &=&-\frac{6-D}{2(D-2)}\left( \frac{D^{\left( 2-D\right) }P_D^2}{2^8\pi ^{D}}\right) ^{1/(4-D)}+  \notag \\
&&+\frac{d^{\left( 2-D\right) }\Gamma (D/2)}{8\left( 2-D\right) \pi ^{D/2}}P_D+O(\delta ).
\label{muStac}
\end{eqnarray}

\noindent For the two dimensional case we have:

\begin{equation}
\mu_2 =\frac{P_2}{16\pi }\ln \left( \frac{e^{\gamma -1/2}}{32\pi }d^{2}P_2\right)
+O(\delta ).
\label{stac3}
\end{equation}

\noindent Note that the integral quantity $W_D=\nu \int_{0}^{d}\theta_D
r^{D-1}dr=P_D\left( 2d^{2}/D-R^{2}\right) /8+O(\delta )$ $\approx
P_Dd^{2}/4D+O(R^{2}/d^{2})$, which is proportional to the power, is also
conserved.

Relations (\ref{stacRva} - \ref{muStac}) completely define the VA
approximate soliton solution in the highly nonlocal case.
It remains to do the same for the AS approximation.

\section{Accessible soliton approximation}

In the AS approximation, the basic assumptions are that the shape of the
nonlocal response of the medium is a parabolic function of the transverse
distance, Eq. (\ref{parabolTheta}), and that the shape function of the field $E$ is still a Gaussian
given by Eq. (\ref{trial1}). The only refractive index "seen" by the
beam is that confined near its propagation axis \cite{snyder}.

The parameters of the trial function Eq. (\ref{trial1}) are given now by equations:

\begin{equation}
\frac{dA}{dz}=-DCA,
\end{equation}

\begin{equation}
C=\frac{1}{2R}\frac{dR}{dz},  \label{as1}
\end{equation}

\begin{equation}
\frac{d^{2}R}{dz^{2}}=\left( \frac{1}{R^{3}}-\Theta_D R\right) ,  \label{as2}
\end{equation}

\begin{equation}
\frac{d\psi }{dz}=-\frac{D}{2R^{2}}+\frac{\theta _{D \max }}{2}.  \label{as3}
\end{equation}

\noindent Equation (\ref{trial1}), together with Eq. (\ref{parabolTheta}), exactly satisfies Eq. (\ref{eq1}).

The parameter $\theta _{D \max}$ is only a phase shift and as such quite
arbitrary in the AS approximation. On the other hand, the value of $\Theta_D $
is much more important; it is determined by Eq. (\ref{theta_D}%
). In this manner, Eq. (\ref{as2}) becomes:

\begin{equation}
\frac{d^{2}R}{dz^{2}}=\frac{1}{R^{3}}-\frac{P_D}{4D\pi ^{D/2}R^{D-1}}.
\label{as4}
\end{equation}

\noindent The equilibrium width $R$ in the AS approximation is:

\begin{equation}
R_{AS}=\left(\frac{ 4D\pi ^{D/2}}{P_D}\right) ^{1/(4-D)},
\end{equation}

\noindent and the amplitude $A$:

\begin{equation}
A_{AS}=\frac{2\sqrt{D}}{R_{AS}^{2}}=2\left(\frac{P_D^2}{16\pi^DD^{D/2}}\right)^{1/(4-D)},
\end{equation}

\noindent

The period of small oscillations of the width perturbation around the
equilibrium can be obtained from (\ref{as4}):

\begin{equation}
\Lambda _{AS}=\frac{2\pi }{\sqrt{4-D}}R_{AS}^{2}=\frac{2\pi }{\sqrt{4-D}}\left(\frac{ 4D\pi ^{D/2}}{P_D}\right) ^{2/(4-D)}.
\end{equation}
It remains to compare these expressions with the ones obtained in the VA approach.
\noindent

\begin{figure}
\begin{minipage}{4cm}
\includegraphics[width=4cm,height=4cm]{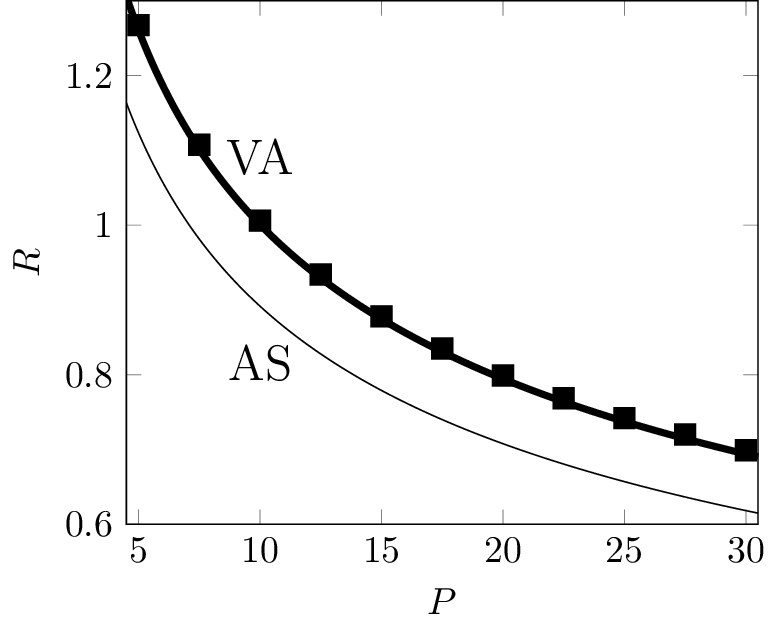}
\subcaption{}
\end{minipage}
\begin{minipage}{4cm}
\includegraphics[width=4cm,height=4cm]{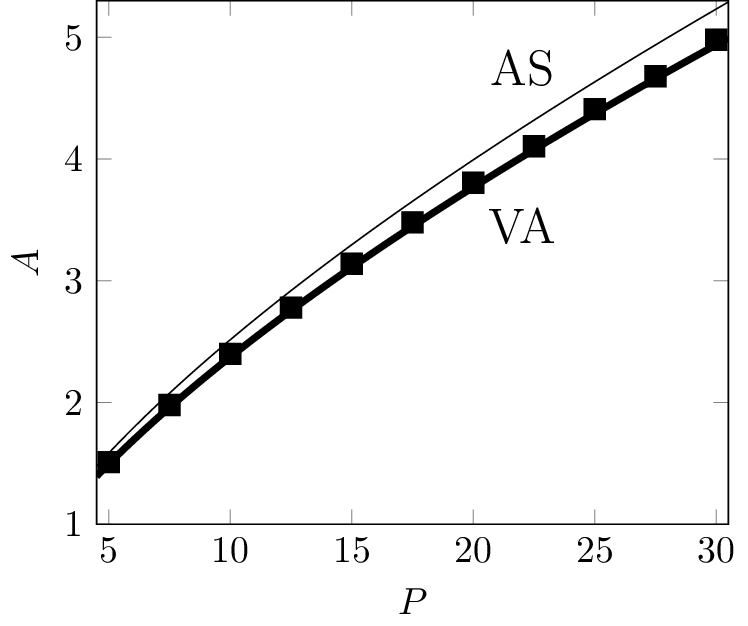}
\subcaption{}
\end{minipage}

\begin{minipage}{4cm}
\includegraphics[width=4cm,height=4cm]{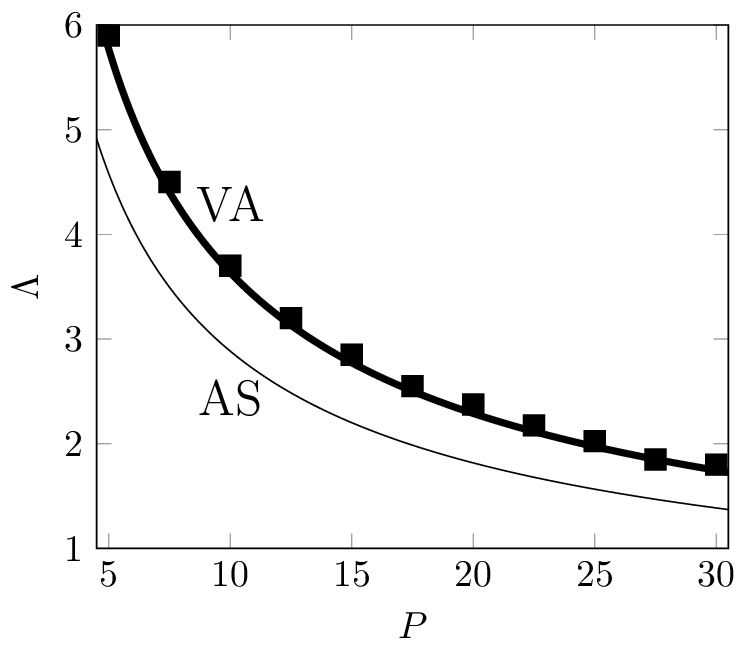}
\subcaption{}
\end{minipage}
\begin{minipage}{4cm}
\includegraphics[width=4cm,height=4cm]{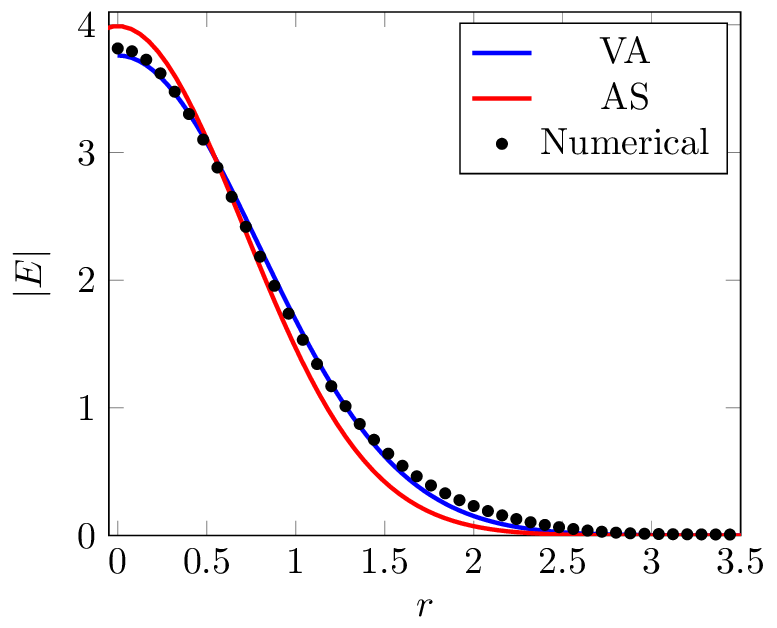}
\subcaption{}
\end{minipage}
\caption{(color online) Comparison of both VA and AS approximations (solid lines) and numerical results (dots) for one dimensional ($D=1$) nonlocal solitons. Beam width $R$ (a),  amplitude $A$ (b), and  period of small oscillations $\Lambda$ (c), are shown as functions of power $P$. (d) Distribution of $|E|$ as a function of $r$ for power $P=20$. Other parameters are $d=40$, $z=100$.
}
\label{1Dpic}
\end{figure}

Another useful approximation in AS approximation is based on the solution of Eq. (\ref{eq2})
in the form of Eq. (\ref{parabolTheta}) when parameter $\Theta_D =A^{2}/\left( 4D\right) =P_D/\left( 4D\pi
^{D/2}R^{D}\right)$ is independent of $z$. In contrast to Eq. (\ref{as4}),
in which the condition $\Theta_D $ may depend on $z$, Eq. (\ref{as2})
for $R$ has an exact oscillatory solution \cite{guo}

\begin{equation}
R=R_{*}\sqrt{\cos ^2(\pi z/\Lambda )+\frac{P_*}{P_D}\sin^2(\pi z/\Lambda )} ,
\end{equation}
from which the solutions for $C$ and $\psi$ immediately follow:

\begin{equation}
C=\frac{\pi }{\Lambda }\frac{\left( P_*/P_D-1\right) \sin (2\pi z/\Lambda )}{
4\left( \cos^2(\pi z/\Lambda )+\left( P_*/P_D\right) \sin ^2(\pi z/\Lambda
)\right) } ,
\end{equation}

\begin{equation}
\psi = - \arctan \left( \sqrt{P_*/P_D}\tan (\pi z/\Lambda )\right) .
\end{equation}

\noindent Here $R_*$ and $P_*$ are the width and the power of the AS
solution, respectively. When $P_D=P_*$, one obtains stationary AS; otherwise,
the approximate solution oscillates. The quantity $\Lambda =\Lambda _{AS}%
\sqrt{P_{*}/P_D}$ represents the period of harmonic oscillations around the
equilibrium (soliton) state, while $\Lambda_{AS}$ is the period of small
oscillations of the width perturbation:

\begin{equation}
\Lambda _{AS}=\pi R_*^2 .
\end{equation}

\noindent Thus, one obtains nice dependencies in closed form, but unfortunately of little
practical value, in view of the large discrepancy with the VA and the numerical
solution of the full problem, to be presented in the next section.

\section{Numerical Results}

For numerical simulations of Eq. (\ref{eq1}) we use finite difference time domain (FDTD)
split-step method \cite{usingGPU}. Equation (\ref{eq2}) is solved by the tridiagonal matrix algorithm (TDMA).
In our simulations we applied the same boundary conditions for both Eqs. (\ref{eq1},\ref{eq2}):

\begin{equation}
\left(\frac{\partial E}{\partial r} \right)_{r=0}=0, \qquad E_{r=d}=0
\end{equation}

and

\begin{equation}
\left(\frac{\partial \theta}{\partial r} \right)_{r=0}=0, \qquad \theta _{r=d}=0
\end{equation}
The initial beam is the trial function given by Eq. (\ref{trial1}),
with parameters determined by VA approximation, Eqs. (\ref{stacRva},\ref{Ava}).

\begin{figure}
\begin{minipage}{4cm}
\includegraphics[width=4cm,height=4cm]{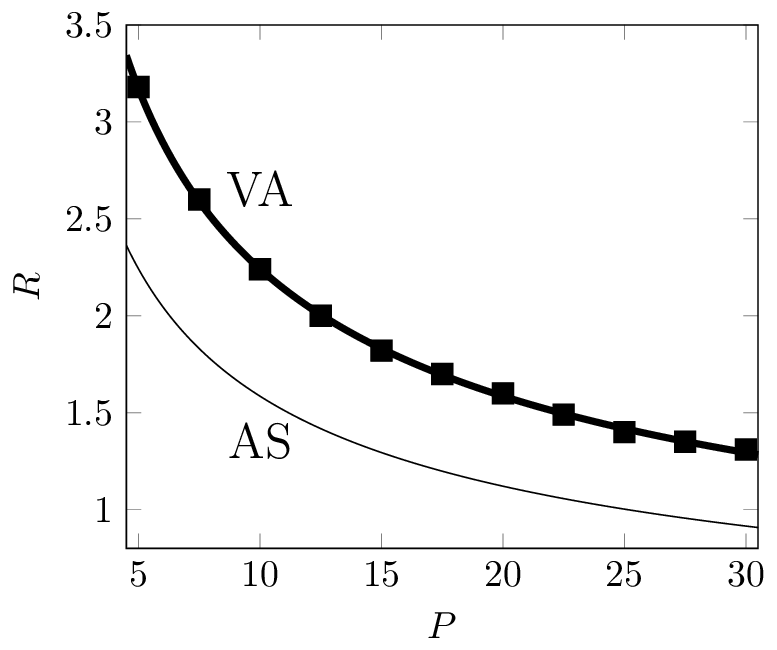}
\subcaption{}
\end{minipage}
\begin{minipage}{4cm}
\includegraphics[width=4cm,height=4cm]{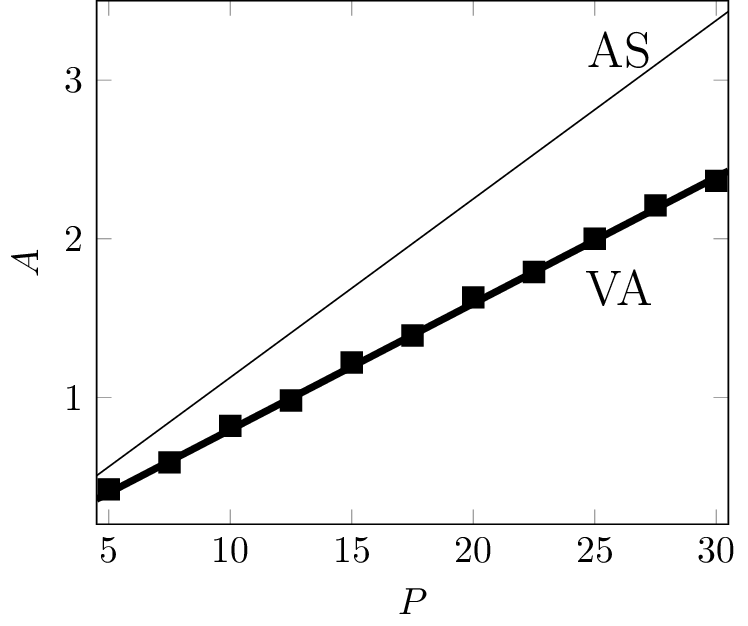}
\subcaption{}
\end{minipage}

\begin{minipage}{4cm}
\includegraphics[width=4cm,height=4cm]{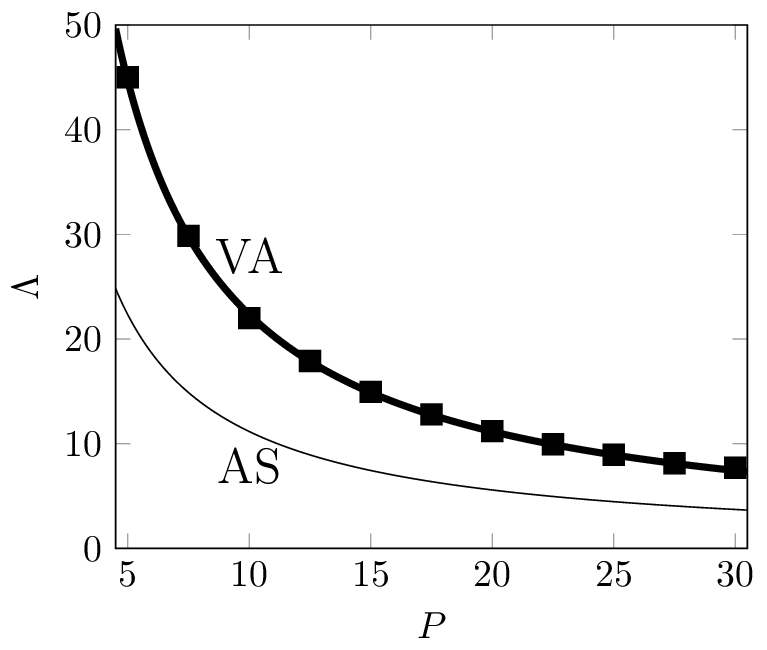}
\subcaption{}
\end{minipage}
\begin{minipage}{4cm}
\includegraphics[width=4cm,height=4cm]{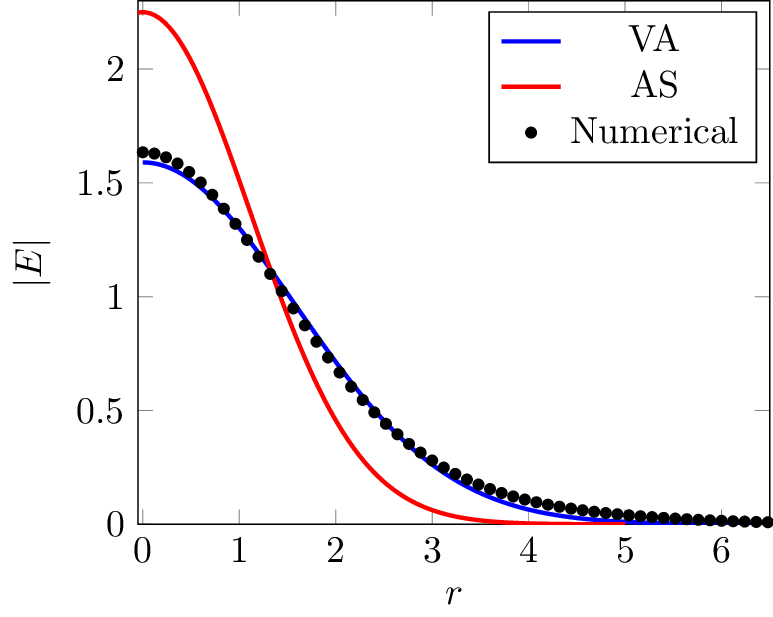}
\subcaption{}
\end{minipage}
\caption{(color online) Same as Fig. (\ref{1Dpic}), but for $D=2$. The only difference is in the propagation length, $z=10^3$.}
\label{2Dpic}
\end{figure}

\begin{figure}
\begin{minipage}{4cm}
\includegraphics[width=4cm,height=4cm]{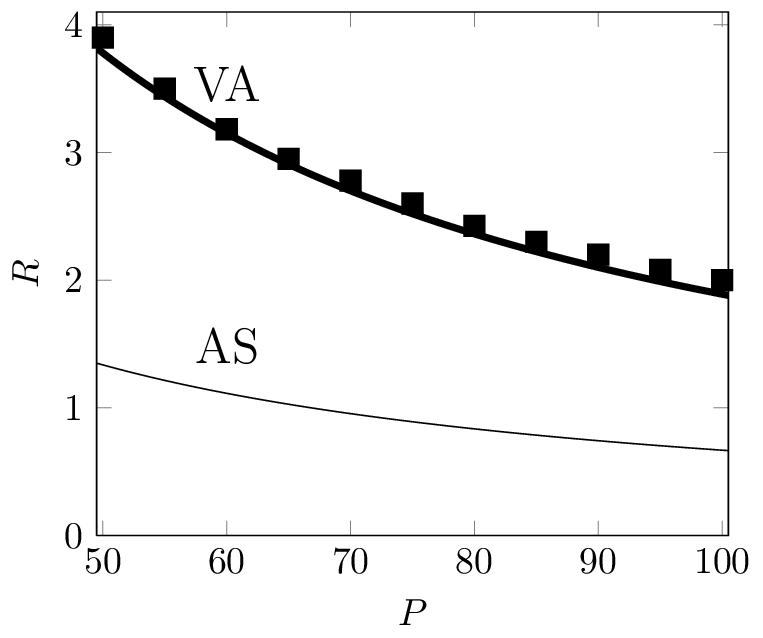}
\subcaption{}
\end{minipage}
\begin{minipage}{4cm}
\includegraphics[width=4cm,height=4cm]{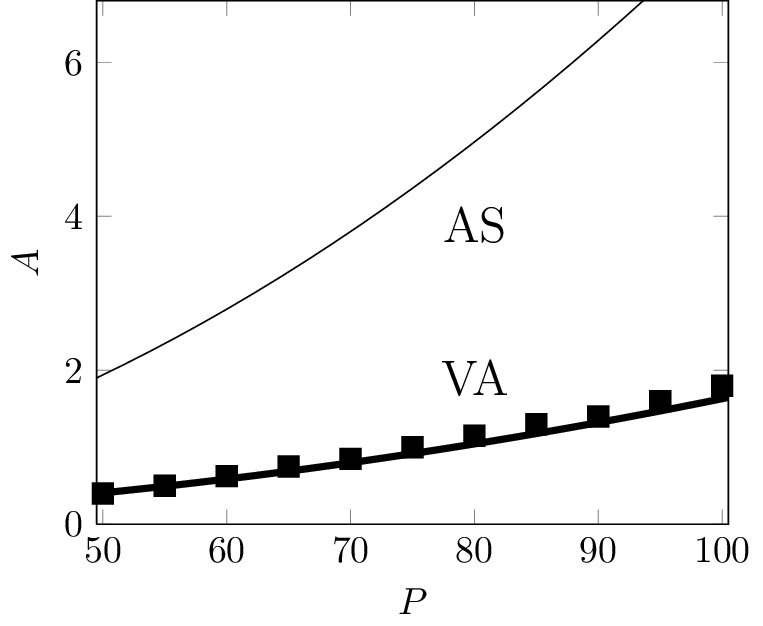}
\subcaption{}
\end{minipage}

\begin{minipage}{4cm}
\includegraphics[width=4cm,height=4cm]{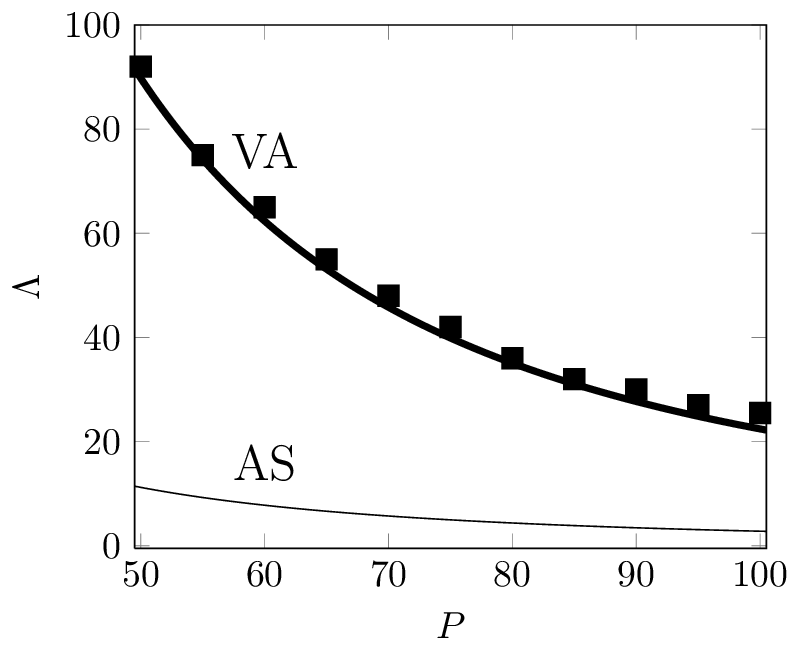}
\subcaption{}
\end{minipage}
\begin{minipage}{4cm}
\includegraphics[width=4cm,height=4cm]{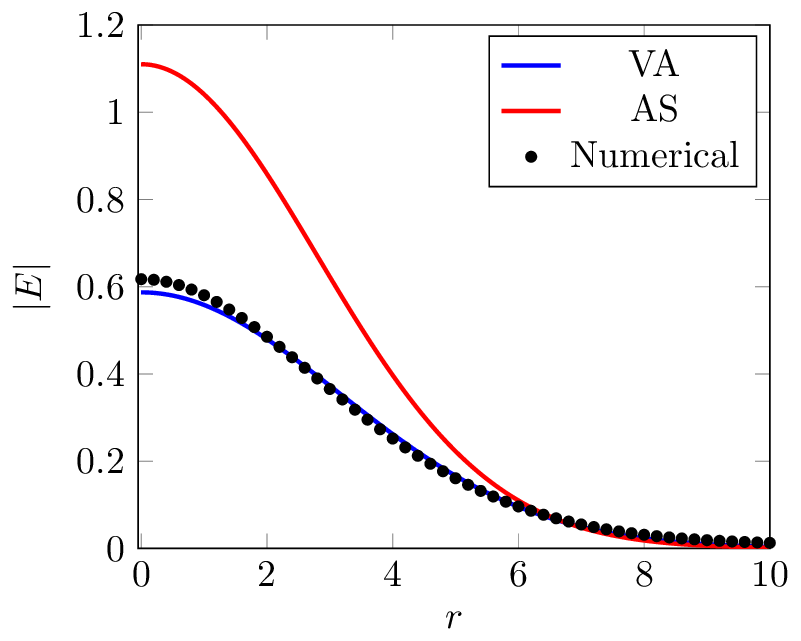}
\subcaption{}
\end{minipage}
\caption{(color online) Same as Fig. (\ref{1Dpic}), but for $D=3$. The propagation distance is now $z=10^4$ and $P=60$.}
\label{3Dpic}
\end{figure}

Figures (\ref{1Dpic},\ref{2Dpic},\ref{3Dpic}) show analytical predictions of both approximations
and numerical results for one, two, and three dimensions, respectively.
The beam width $R$ of soliton solutions for AS approximation is $2^{D/(8-2D)}$ times smaller
than the VA approximation at a same power, see panel (a) in Figs. (\ref{1Dpic},\ref{2Dpic},\ref{3Dpic}).
The corresponding stationary amplitudes in both approximations are presented in panel (b)
of Figs.(\ref{1Dpic},\ref{2Dpic},\ref{3Dpic}). Because $P_D=\pi ^{D/2}A^{2}R^{D}$, the equilibrium amplitude $A$ in the AS approximation is
$2^{D^2/(16-4D)}$ times greater than the one in the VA approximation at the same power.
The period in the AS approximation is $2^{D/(4-D)}$ times less than that in the VA
approximation at the same power, see panel (c) in Figs. (\ref{1Dpic},\ref{2Dpic},\ref{3Dpic}).
Panel (d) in Figs. (\ref{1Dpic},\ref{2Dpic},\ref{3Dpic}) shows an example of the soliton shape
predicted by VA and AS, as well as the numerical output after a long propagation distance.

Thus, the values of the beam parameters for AS are systematically off the values for the VA approximation.
In all the cases the VA predictions are confirmed by numerical results.
This confirms that, while AS approximation may represent a valuable aid for an easy understanding of
highly nonlocal solitons, it is a poor approximation to the more exact approaches, such as the VA approximation,
and to the numerical solution.

\section{Conclusion}

In conclusion, we have discussed the differences between the VA and AS
approximate solutions to the propagation of solitons in highly nonlocal
nonlinear media. The AS model provides a radical simplification of the problem and allows
for an elegant description of solitons, but has a limited practical relevance, mainly
because of the competition between the nonlocality and the finite size of
the sample, which leads to systematic errors. The VA solution is not so simple, but it works very well in the
limited region of large nonlocality. We have found that the AS approximation
can differ up to eight times, when compared to the more realistic VA
approximation and the numerical solution.

\begin{acknowledgments}
This publication was made possible by NPRP Grants  \# 5 - 674 - 1 -114 and \# 6-021-1-005 from the 
Qatar National Research Fund (a member of the Qatar Foundation). The statements made herein are 
solely the responsibility of the authors. Work at the Institute of Physics Belgrade was 
supported by the Ministry of Science of the Republic of Serbia under the projects No. OI 171033, No. 171006 and III 46016.
\end{acknowledgments}

\bibliographystyle{apsrev4-1}

\bibliography{reference}

\begin{thebibliography}{19}%
\makeatletter
\providecommand \@ifxundefined [1]{%
 \@ifx{#1\undefined}
}%
\providecommand \@ifnum [1]{%
 \ifnum #1\expandafter \@firstoftwo
 \else \expandafter \@secondoftwo
 \fi
}%
\providecommand \@ifx [1]{%
 \ifx #1\expandafter \@firstoftwo
 \else \expandafter \@secondoftwo
 \fi
}%
\providecommand \natexlab [1]{#1}%
\providecommand \enquote  [1]{``#1''}%
\providecommand \bibnamefont  [1]{#1}%
\providecommand \bibfnamefont [1]{#1}%
\providecommand \citenamefont [1]{#1}%
\providecommand \href@noop [0]{\@secondoftwo}%
\providecommand \href [0]{\begingroup \@sanitize@url \@href}%
\providecommand \@href[1]{\@@startlink{#1}\@@href}%
\providecommand \@@href[1]{\endgroup#1\@@endlink}%
\providecommand \@sanitize@url [0]{\catcode `\\12\catcode `\$12\catcode
  `\&12\catcode `\#12\catcode `\^12\catcode `\_12\catcode `\%12\relax}%
\providecommand \@@startlink[1]{}%
\providecommand \@@endlink[0]{}%
\providecommand \url  [0]{\begingroup\@sanitize@url \@url }%
\providecommand \@url [1]{\endgroup\@href {#1}{\urlprefix }}%
\providecommand \urlprefix  [0]{URL }%
\providecommand \Eprint [0]{\href }%
\providecommand \doibase [0]{http://dx.doi.org/}%
\providecommand \selectlanguage [0]{\@gobble}%
\providecommand \bibinfo  [0]{\@secondoftwo}%
\providecommand \bibfield  [0]{\@secondoftwo}%
\providecommand \translation [1]{[#1]}%
\providecommand \BibitemOpen [0]{}%
\providecommand \bibitemStop [0]{}%
\providecommand \bibitemNoStop [0]{.\EOS\space}%
\providecommand \EOS [0]{\spacefactor3000\relax}%
\providecommand \BibitemShut  [1]{\csname bibitem#1\endcsname}%
\let\auto@bib@innerbib\@empty
\bibitem [{\citenamefont {Kivshar}\ and\ \citenamefont {Agrawal}(2003)}]{yuri}%
  \BibitemOpen
  \bibfield  {author} {\bibinfo {author} {\bibfnamefont {Y.~S.}\ \bibnamefont
  {Kivshar}}\ and\ \bibinfo {author} {\bibfnamefont {G.}~\bibnamefont
  {Agrawal}},\ }\href@noop {} {\emph {\bibinfo {title} {Optical solitons: from
  fibers to photonic crystals}}}\ (\bibinfo  {publisher} {Academic press},\
  \bibinfo {year} {2003})\BibitemShut {NoStop}%
\bibitem [{\citenamefont {Henninot}\ \emph {et~al.}(2007)\citenamefont
  {Henninot}, \citenamefont {Blach},\ and\ \citenamefont {Warenghem}}]{hen}%
  \BibitemOpen
  \bibfield  {author} {\bibinfo {author} {\bibfnamefont {J.}~\bibnamefont
  {Henninot}}, \bibinfo {author} {\bibfnamefont {J.}~\bibnamefont {Blach}}, \
  and\ \bibinfo {author} {\bibfnamefont {M.}~\bibnamefont {Warenghem}},\
  }\href@noop {} {\bibfield  {journal} {\bibinfo  {journal} {J. Opt. A: Pure
  and Applied Opt.}\ }\textbf {\bibinfo {volume} {9}},\ \bibinfo {pages} {20}
  (\bibinfo {year} {2007})}\BibitemShut {NoStop}%
\bibitem [{\citenamefont {Hutsebaut}\ \emph {et~al.}(2005)\citenamefont
  {Hutsebaut}, \citenamefont {Cambournac}, \citenamefont {Haelterman},
  \citenamefont {Beeckman},\ and\ \citenamefont {Neyts}}]{cyril}%
  \BibitemOpen
  \bibfield  {author} {\bibinfo {author} {\bibfnamefont {X.}~\bibnamefont
  {Hutsebaut}}, \bibinfo {author} {\bibfnamefont {C.}~\bibnamefont
  {Cambournac}}, \bibinfo {author} {\bibfnamefont {M.}~\bibnamefont
  {Haelterman}}, \bibinfo {author} {\bibfnamefont {J.}~\bibnamefont
  {Beeckman}}, \ and\ \bibinfo {author} {\bibfnamefont {K.}~\bibnamefont
  {Neyts}},\ }\href {\doibase 10.1364/JOSAB.22.001424} {\bibfield  {journal}
  {\bibinfo  {journal} {J. Opt. Soc. Am. B}\ }\textbf {\bibinfo {volume}
  {22}},\ \bibinfo {pages} {1424} (\bibinfo {year} {2005})}\BibitemShut
  {NoStop}%
\bibitem [{\citenamefont {Beeckman}\ \emph {et~al.}(2004)\citenamefont
  {Beeckman}, \citenamefont {Neyts}, \citenamefont {Hutsebaut}, \citenamefont
  {Cambournac},\ and\ \citenamefont {Haelterman}}]{beec}%
  \BibitemOpen
  \bibfield  {author} {\bibinfo {author} {\bibfnamefont {J.}~\bibnamefont
  {Beeckman}}, \bibinfo {author} {\bibfnamefont {K.}~\bibnamefont {Neyts}},
  \bibinfo {author} {\bibfnamefont {X.}~\bibnamefont {Hutsebaut}}, \bibinfo
  {author} {\bibfnamefont {C.}~\bibnamefont {Cambournac}}, \ and\ \bibinfo
  {author} {\bibfnamefont {M.}~\bibnamefont {Haelterman}},\ }\href@noop {}
  {\bibfield  {journal} {\bibinfo  {journal} {Opt. Express}\ }\textbf {\bibinfo
  {volume} {12}},\ \bibinfo {pages} {1011} (\bibinfo {year}
  {2004})}\BibitemShut {NoStop}%
\bibitem [{\citenamefont {Snyder}\ and\ \citenamefont
  {Mitchell}(1997)}]{snyder}%
  \BibitemOpen
  \bibfield  {author} {\bibinfo {author} {\bibfnamefont {A.~W.}\ \bibnamefont
  {Snyder}}\ and\ \bibinfo {author} {\bibfnamefont {D.~J.}\ \bibnamefont
  {Mitchell}},\ }\href@noop {} {\bibfield  {journal} {\bibinfo  {journal}
  {Science}\ }\textbf {\bibinfo {volume} {276}},\ \bibinfo {pages} {1538}
  (\bibinfo {year} {1997})}\BibitemShut {NoStop}%
\bibitem [{\citenamefont {Conti}\ \emph {et~al.}(2003)\citenamefont {Conti},
  \citenamefont {Peccianti},\ and\ \citenamefont {Assanto}}]{assanto2003}%
  \BibitemOpen
  \bibfield  {author} {\bibinfo {author} {\bibfnamefont {C.}~\bibnamefont
  {Conti}}, \bibinfo {author} {\bibfnamefont {M.}~\bibnamefont {Peccianti}}, \
  and\ \bibinfo {author} {\bibfnamefont {G.}~\bibnamefont {Assanto}},\
  }\href@noop {} {\bibfield  {journal} {\bibinfo  {journal} {Phys. Rev. Lett.}\
  }\textbf {\bibinfo {volume} {91}},\ \bibinfo {pages} {073901} (\bibinfo
  {year} {2003})}\BibitemShut {NoStop}%
\bibitem [{\citenamefont {Conti}\ \emph {et~al.}(2004)\citenamefont {Conti},
  \citenamefont {Peccianti},\ and\ \citenamefont {Assanto}}]{assanto2004}%
  \BibitemOpen
  \bibfield  {author} {\bibinfo {author} {\bibfnamefont {C.}~\bibnamefont
  {Conti}}, \bibinfo {author} {\bibfnamefont {M.}~\bibnamefont {Peccianti}}, \
  and\ \bibinfo {author} {\bibfnamefont {G.}~\bibnamefont {Assanto}},\
  }\href@noop {} {\bibfield  {journal} {\bibinfo  {journal} {Phys. Rev. Lett.}\
  }\textbf {\bibinfo {volume} {92}},\ \bibinfo {pages} {113902} (\bibinfo
  {year} {2004})}\BibitemShut {NoStop}%
\bibitem [{\citenamefont {Henninot}\ \emph {et~al.}(2008)\citenamefont
  {Henninot}, \citenamefont {Blach},\ and\ \citenamefont
  {Warenghem}}]{henninot}%
  \BibitemOpen
  \bibfield  {author} {\bibinfo {author} {\bibfnamefont {J.}~\bibnamefont
  {Henninot}}, \bibinfo {author} {\bibfnamefont {J.}~\bibnamefont {Blach}}, \
  and\ \bibinfo {author} {\bibfnamefont {M.}~\bibnamefont {Warenghem}},\
  }\href@noop {} {\bibfield  {journal} {\bibinfo  {journal} {Journal of Optics
  A: Pure and Applied Optics}\ }\textbf {\bibinfo {volume} {10}},\ \bibinfo
  {pages} {085104} (\bibinfo {year} {2008})}\BibitemShut {NoStop}%
\bibitem [{\citenamefont {Aleksi{\'c}}\ \emph {et~al.}(2012)\citenamefont
  {Aleksi{\'c}}, \citenamefont {Petrovi{\'c}}, \citenamefont {Strini{\'c}},\
  and\ \citenamefont {Beli{\'c}}}]{aleksic}%
  \BibitemOpen
  \bibfield  {author} {\bibinfo {author} {\bibfnamefont {N.~B.}\ \bibnamefont
  {Aleksi{\'c}}}, \bibinfo {author} {\bibfnamefont {M.~S.}\ \bibnamefont
  {Petrovi{\'c}}}, \bibinfo {author} {\bibfnamefont {A.~I.}\ \bibnamefont
  {Strini{\'c}}}, \ and\ \bibinfo {author} {\bibfnamefont {M.~R.}\ \bibnamefont
  {Beli{\'c}}},\ }\href@noop {} {\bibfield  {journal} {\bibinfo  {journal}
  {Phys. Rev. A}\ }\textbf {\bibinfo {volume} {85}},\ \bibinfo {pages} {033826}
  (\bibinfo {year} {2012})}\BibitemShut {NoStop}%
\bibitem [{\citenamefont {Buccoliero}\ \emph {et~al.}(2009)\citenamefont
  {Buccoliero}, \citenamefont {Desyatnikov}, \citenamefont {Krolikowski},\ and\
  \citenamefont {Kivshar}}]{bucco}%
  \BibitemOpen
  \bibfield  {author} {\bibinfo {author} {\bibfnamefont {D.}~\bibnamefont
  {Buccoliero}}, \bibinfo {author} {\bibfnamefont {A.~S.}\ \bibnamefont
  {Desyatnikov}}, \bibinfo {author} {\bibfnamefont {W.}~\bibnamefont
  {Krolikowski}}, \ and\ \bibinfo {author} {\bibfnamefont {Y.~S.}\ \bibnamefont
  {Kivshar}},\ }\href@noop {} {\bibfield  {journal} {\bibinfo  {journal}
  {Journal of Optics A: Pure and Applied Optics}\ }\textbf {\bibinfo {volume}
  {11}},\ \bibinfo {pages} {094014} (\bibinfo {year} {2009})}\BibitemShut
  {NoStop}%
\bibitem [{\citenamefont {Beli\ifmmode~\acute{c}\else \'{c}\fi{}}\ \emph
  {et~al.}(2002)\citenamefont {Beli\ifmmode~\acute{c}\else \'{c}\fi{}},
  \citenamefont {Vuji\ifmmode~\acute{c}\else \'{c}\fi{}}, \citenamefont
  {Stepken}, \citenamefont {Kaiser}, \citenamefont {Calvo}, \citenamefont
  {Agull\'o-L\'opez},\ and\ \citenamefont {Carrascosa}}]{belic}%
  \BibitemOpen
  \bibfield  {author} {\bibinfo {author} {\bibfnamefont {M.~R.}\ \bibnamefont
  {Beli\ifmmode~\acute{c}\else \'{c}\fi{}}}, \bibinfo {author} {\bibfnamefont
  {D.}~\bibnamefont {Vuji\ifmmode~\acute{c}\else \'{c}\fi{}}}, \bibinfo
  {author} {\bibfnamefont {A.}~\bibnamefont {Stepken}}, \bibinfo {author}
  {\bibfnamefont {F.}~\bibnamefont {Kaiser}}, \bibinfo {author} {\bibfnamefont
  {G.~F.}\ \bibnamefont {Calvo}}, \bibinfo {author} {\bibfnamefont
  {F.}~\bibnamefont {Agull\'o-L\'opez}}, \ and\ \bibinfo {author}
  {\bibfnamefont {M.}~\bibnamefont {Carrascosa}},\ }\href {\doibase
  10.1103/PhysRevE.65.066610} {\bibfield  {journal} {\bibinfo  {journal} {Phys.
  Rev. E}\ }\textbf {\bibinfo {volume} {65}},\ \bibinfo {pages} {066610}
  (\bibinfo {year} {2002})}\BibitemShut {NoStop}%
\bibitem [{\citenamefont {Santos}\ \emph {et~al.}(2003)\citenamefont {Santos},
  \citenamefont {Shlyapnikov},\ and\ \citenamefont {Lewenstein}}]{bec}%
  \BibitemOpen
  \bibfield  {author} {\bibinfo {author} {\bibfnamefont {L.}~\bibnamefont
  {Santos}}, \bibinfo {author} {\bibfnamefont {G.~V.}\ \bibnamefont
  {Shlyapnikov}}, \ and\ \bibinfo {author} {\bibfnamefont {M.}~\bibnamefont
  {Lewenstein}},\ }\href {\doibase 10.1103/PhysRevLett.90.250403} {\bibfield
  {journal} {\bibinfo  {journal} {Phys. Rev. Lett.}\ }\textbf {\bibinfo
  {volume} {90}},\ \bibinfo {pages} {250403} (\bibinfo {year}
  {2003})}\BibitemShut {NoStop}%
\bibitem [{\citenamefont {Petrovi{\'c}}\ \emph {et~al.}(2013)\citenamefont
  {Petrovi{\'c}}, \citenamefont {Aleksi{\'c}}, \citenamefont {Strini{\'c}},\
  and\ \citenamefont {Beli{\'c}}}]{petrovic}%
  \BibitemOpen
  \bibfield  {author} {\bibinfo {author} {\bibfnamefont {M.~S.}\ \bibnamefont
  {Petrovi{\'c}}}, \bibinfo {author} {\bibfnamefont {N.~B.}\ \bibnamefont
  {Aleksi{\'c}}}, \bibinfo {author} {\bibfnamefont {A.~I.}\ \bibnamefont
  {Strini{\'c}}}, \ and\ \bibinfo {author} {\bibfnamefont {M.~R.}\ \bibnamefont
  {Beli{\'c}}},\ }\href@noop {} {\bibfield  {journal} {\bibinfo  {journal}
  {Phys. Rev. A}\ }\textbf {\bibinfo {volume} {87}},\ \bibinfo {pages} {043825}
  (\bibinfo {year} {2013})}\BibitemShut {NoStop}%
\bibitem [{\citenamefont {Assanto}\ and\ \citenamefont
  {Smyth}(2013)}]{comment}%
  \BibitemOpen
  \bibfield  {author} {\bibinfo {author} {\bibfnamefont {G.}~\bibnamefont
  {Assanto}}\ and\ \bibinfo {author} {\bibfnamefont {N.~F.}\ \bibnamefont
  {Smyth}},\ }\href@noop {} {\bibfield  {journal} {\bibinfo  {journal} {Phys.
  Rev. A}\ }\textbf {\bibinfo {volume} {87}},\ \bibinfo {pages} {047801}
  (\bibinfo {year} {2013})}\BibitemShut {NoStop}%
\bibitem [{\citenamefont {Aleksi\ifmmode~\acute{c}\else \'{c}\fi{}}\ \emph
  {et~al.}(2013)\citenamefont {Aleksi\ifmmode~\acute{c}\else \'{c}\fi{}},
  \citenamefont {Petrovi\ifmmode~\acute{c}\else \'{c}\fi{}}, \citenamefont
  {Strini\ifmmode~\acute{c}\else \'{c}\fi{}},\ and\ \citenamefont
  {Beli\ifmmode~\acute{c}\else \'{c}\fi{}}}]{reply}%
  \BibitemOpen
  \bibfield  {author} {\bibinfo {author} {\bibfnamefont {N.~B.}\ \bibnamefont
  {Aleksi\ifmmode~\acute{c}\else \'{c}\fi{}}}, \bibinfo {author} {\bibfnamefont
  {M.~S.}\ \bibnamefont {Petrovi\ifmmode~\acute{c}\else \'{c}\fi{}}}, \bibinfo
  {author} {\bibfnamefont {A.~I.}\ \bibnamefont {Strini\ifmmode~\acute{c}\else
  \'{c}\fi{}}}, \ and\ \bibinfo {author} {\bibfnamefont {M.~R.}\ \bibnamefont
  {Beli\ifmmode~\acute{c}\else \'{c}\fi{}}},\ }\href {\doibase
  10.1103/PhysRevA.87.047802} {\bibfield  {journal} {\bibinfo  {journal} {Phys.
  Rev. A}\ }\textbf {\bibinfo {volume} {87}},\ \bibinfo {pages} {047802}
  (\bibinfo {year} {2013})}\BibitemShut {NoStop}%
\bibitem [{\citenamefont {Abramowitz}\ and\ \citenamefont
  {Stegun}(1964)}]{abramowitz}%
  \BibitemOpen
  \bibfield  {author} {\bibinfo {author} {\bibfnamefont {M.}~\bibnamefont
  {Abramowitz}}\ and\ \bibinfo {author} {\bibfnamefont {I.~A.}\ \bibnamefont
  {Stegun}},\ }\href@noop {} {\emph {\bibinfo {title} {Handbook of Mathematical
  Functions: With Formulars, Graphs, and Mathematical Tables}}},\ Vol.~\bibinfo
  {volume} {55}\ (\bibinfo  {publisher} {DoverPublications. com},\ \bibinfo
  {year} {1964})\BibitemShut {NoStop}%
\bibitem [{\citenamefont {Aleksi{\'c}}\ \emph {et~al.}(2013)\citenamefont
  {Aleksi{\'c}}, \citenamefont {Aleksi{\'c}}, \citenamefont {Petrovi{\'c}},
  \citenamefont {Strini{\'c}},\ and\ \citenamefont {Beli{\'c}}}]{VAvsAS}%
  \BibitemOpen
  \bibfield  {author} {\bibinfo {author} {\bibfnamefont {B.}~\bibnamefont
  {Aleksi{\'c}}}, \bibinfo {author} {\bibfnamefont {N.}~\bibnamefont
  {Aleksi{\'c}}}, \bibinfo {author} {\bibfnamefont {M.~S.}\ \bibnamefont
  {Petrovi{\'c}}}, \bibinfo {author} {\bibfnamefont {A.~I.}\ \bibnamefont
  {Strini{\'c}}}, \ and\ \bibinfo {author} {\bibfnamefont {M.~R.}\ \bibnamefont
  {Beli{\'c}}},\ }\href@noop {} {\bibfield  {journal} {\bibinfo  {journal}
  {arXiv preprint arXiv:1311.6840}\ } (\bibinfo {year} {2013})}\BibitemShut
  {NoStop}%
\bibitem [{\citenamefont {Guo}\ \emph {et~al.}(2004)\citenamefont {Guo},
  \citenamefont {Luo}, \citenamefont {Yi}, \citenamefont {Chi},\ and\
  \citenamefont {Xie}}]{guo}%
  \BibitemOpen
  \bibfield  {author} {\bibinfo {author} {\bibfnamefont {Q.}~\bibnamefont
  {Guo}}, \bibinfo {author} {\bibfnamefont {B.}~\bibnamefont {Luo}}, \bibinfo
  {author} {\bibfnamefont {F.}~\bibnamefont {Yi}}, \bibinfo {author}
  {\bibfnamefont {S.}~\bibnamefont {Chi}}, \ and\ \bibinfo {author}
  {\bibfnamefont {Y.}~\bibnamefont {Xie}},\ }\href@noop {} {\bibfield
  {journal} {\bibinfo  {journal} {Phys. Rev. E}\ }\textbf {\bibinfo {volume}
  {69}},\ \bibinfo {pages} {016602} (\bibinfo {year} {2004})}\BibitemShut
  {NoStop}%
\bibitem [{\citenamefont {Aleksic}\ \emph {et~al.}(2012)\citenamefont
  {Aleksic}, \citenamefont {Aleksic}, \citenamefont {Skarka},\ and\
  \citenamefont {Belic}}]{usingGPU}%
  \BibitemOpen
  \bibfield  {author} {\bibinfo {author} {\bibfnamefont {B.}~\bibnamefont
  {Aleksic}}, \bibinfo {author} {\bibfnamefont {N.}~\bibnamefont {Aleksic}},
  \bibinfo {author} {\bibfnamefont {V.}~\bibnamefont {Skarka}}, \ and\ \bibinfo
  {author} {\bibfnamefont {M.}~\bibnamefont {Belic}},\ }\href
  {http://stacks.iop.org/1402-4896/2012/i=T149/a=014036} {\bibfield  {journal}
  {\bibinfo  {journal} {Physica Scripta}\ }\textbf {\bibinfo {volume} {2012}},\
  \bibinfo {pages} {014036} (\bibinfo {year} {2012})}\BibitemShut {NoStop}%
\end{thebibliography}%

\end{document}